\documentclass[12pt,a4paper,final]{iopart}
\usepackage{iopams}  
\usepackage[breaklinks=true,colorlinks=true,linkcolor=blue,urlcolor=blue,citecolor=blue]{hyperref}
\usepackage{amsfonts}%
\usepackage{amssymb}%
\usepackage{graphicx}
\usepackage{color}
\usepackage{tabularx}
\usepackage{braket}
\begin{document}

\title[The 3D Split-Ring Cavity Lattice]{The 3D Split-Ring Cavity Lattice: A New Metastructure for Engineering Arrays of Coupled Microwave Harmonic Oscillators}

\author[cor1]{Maxim Goryachev$^1$}
\address{$^1$ARC Centre of Excellence for Engineered Quantum Systems, University of Western Australia, 35 Stirling Highway, Crawley WA 6009, Australia}
\ead{\mailto{maxim.goryachev@uwa.edu.au}}

\author{Michael E. Tobar$^1$}
\address{$^1$ARC Centre of Excellence for Engineered Quantum Systems, University of Western Australia, 35 Stirling Highway, Crawley WA 6009, Australia}
\ead{michael.tobar@uwa.edu.au}

\date{\today}


\begin{abstract}

A new electromagnetic cavity structure, a lattice of 3D cavities consisting of an array of posts and gaps is presented. The individual cavity elements are based on the cylindrical re-entrant (or Klystron) cavity. We show that these cavities can also can be thought of as 3D split-ring resonators, which is confirmed by applying symmetry transformations, each of which is an electromagnetic resonator with spatially separated magnetic and electric field. The characteristics of the cavity is used to mimic phonon behaviour of a one dimensional chain of atoms. It is demonstrated how magnetic field coupling can lead to phonon-like dispersion curves with acoustical and optical branches. The system is able to reproduce a number of effects typical to one-dimensional lattices exhibiting acoustic vibration, such as band gaps, phonon trapping, and effects of impurities. In addition, quasicrystal emulations predict the results expected from this class of ordered structures. The system is easily scalable to simulate 2D and 3D lattices and shows a new way to engineer arrays of coupled microwave resonators with a variety of possible applications to hybrid quantum systems proposed.

\end{abstract}

\submitto{\NJP}

\maketitle

\section*{Introduction}

Metamaterials are structures made of artificial building blocks with a scale smaller than the working wavelength\cite{Zheludev:2010li}. Such materials are typically designed to possess properties that cannot be found in natural materials. The classical example of such an application is metamaterials with negative refractive index\cite{Valentine:2008fz}. Such metamaterials are definitely in need in both science and engineering. Although, there could be another potential application of metamaterials: They can be used to emulate the behaviour of systems from one physical realm by making experiments in another one, potentially with some boosted parameters. For example, can the behaviour of a mechanical solid be emulated in an electromagnetic cavity with the speed of sound approaching the speed of light? Such systems can also be used to study properties of solids that cannot be found in nature, for example quasicrystals or phonon trapping lattices.

Although the answer to the question asked above is definitely yes, it is important to propose a practical structure that is simple, scalable, tuneable and with low loss. In this work, we analyse a new type of metamaterial based on the cylindrical re-entrant (or Klystron) cavity, which is the 3D analog of the 2D split-ring resonator with similar characteristics. 2D split-ring resonators have been a popular choice for metamaterials\cite{Smith:2000fk,Ates:2010kx}. However,  one of the weak points of the 2D structure is the inability to reach very high quality factors. Thus, there is an interest to explore metamaterial made of 3D cavities where all the benefits could be fully exploited. In this work we take this approach and undertake 3D finite-element electromagnetic modelling of a multiple post 3D cavities and compare the results to the phonon behaviour in a one  dimensional chain of atoms. These similarities enable the emulation of defects in crystals, as well as the emulation of some properties of quasicrystals.

\section{System Description}

\begin{figure}[t!]
	\centering
			\includegraphics[width=0.15\textwidth]{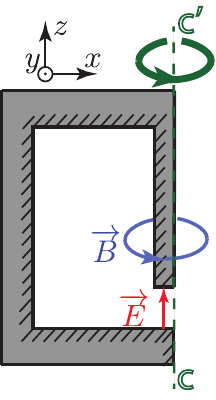}
	\caption{Rotation of the split-ring resonator (a surface) required to construct a 3D microwave cavity (a solid).}
	\label{splitring}
\end{figure}

The one dimensional metamaterial analysed in this work is based on the re-enrant cavity { (also sometimes referred to as a Klystron cavity\cite{reen0,reen1,reen2}).} The re-entrant cavity is a closed (typically cylindrical) 3D microwave resonator built around a central post with a tiny gap between one of the cavity walls and the post tip. It can be demonstrated that this cavity can be considered as a continuous $2\pi$ rotation of a 2D split-ring resonator by continuously rotating a two dimensional surface, a split-ring resonator, around axis of rotation $C^\prime C$ by $2\pi$ in the $yx$ plane (shown in Fig.~\ref{splitring}). As a result such cavities demonstrates continuous rotational symmetry.

Unlike most other 3D structures the re-entrant cavities have spatially separated electrical and magnetic fields, with almost all the electrical field concentrated in the gap and all of the magnetic field distributed around the post, which decays rapidly with the distance from it. Thus, as a resonant system, the cavity can be approximated by an $LC$ model in the vicinity of its main resonance. { Expressions for equivalent inductance and capacitance of a single-post re-entrant cavity have been calculated previously\cite{reen0}. These relations approximate the equivalent capacitance $C$ by a capacitor formed between the post tip and the opposite cavity surface $C = \frac{\pi\varepsilon r^2}{d}$, where $d$ is the gap size, $r$ is the post radius and the $\varepsilon$ is the permittivity. In addition, for some extreme values of the post relative to the cavity size, some other parasitic capacitance may become significant. The equivalent inductance is related to both cavity $R$ and post $r$ radii, as well as the post length and is given by: $L = \frac{h\mu_0}{2\pi}\ln\frac{R}{r}$.}

{ One advantage of the 3D matematerial geometry over its 2D analog, is the better field confinement within the resonator (no leakage). This can lead to higher quality factors, which has already been demonstrated for single post re-entrant cavity transducers designed for gravitational wave detection\cite{Bassan2008}. In order to couple to the confined field within the cavity, standard techniques such as coaxial cables with antenna or loop probes may be implemented. Such cables are inserted into the inner part of the cavity through a hole\cite{reen0}. If the probe is under the post, the straight antenna is preferred, otherwise a loop probe from the side, which couples to the circulating magnetic field may be implemented.}

\begin{figure}[t!]
	\centering
			\includegraphics[width=0.65\textwidth]{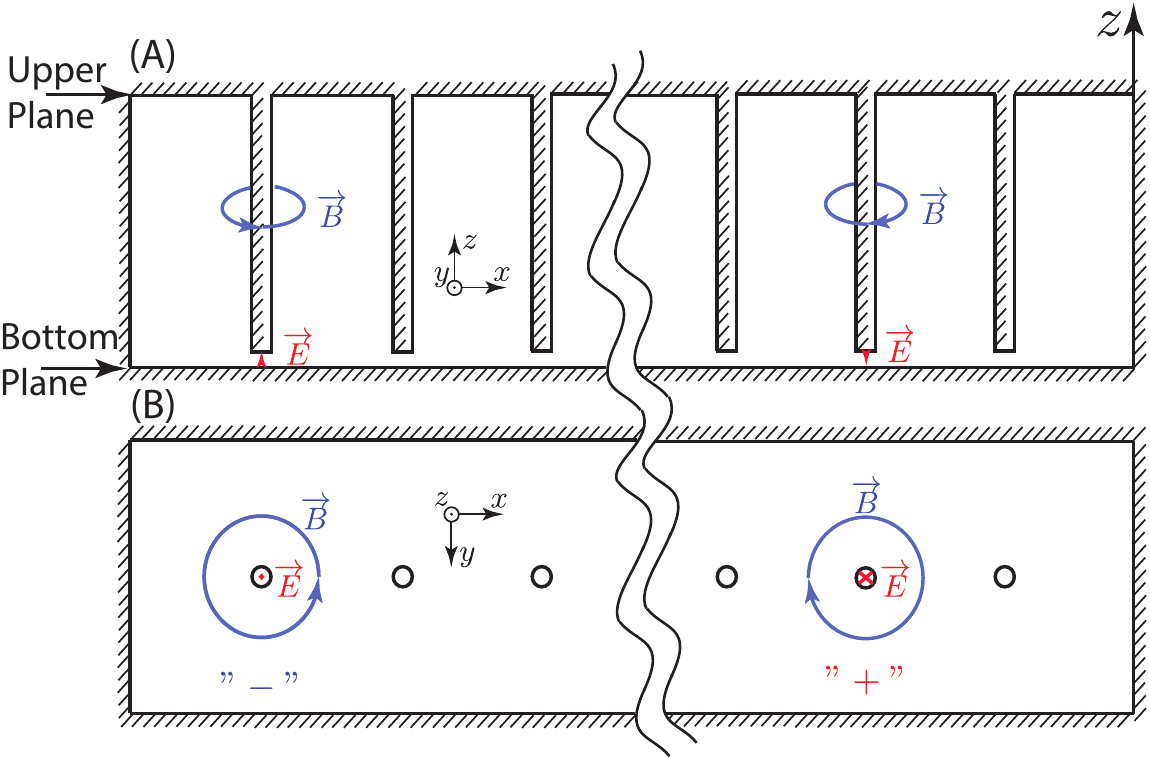}
	\caption{(A) Side and (B) Top views of a metamaterial lattice of the re-entrant-type cavity resonance posts constituting an 1D lattice. Confinement of magnetic and electrical fields are shown only for two posts demonstrating two possible orientations.}
	\label{meta1}
\end{figure}

The metamaterial analysed in this work is constructed from a standard single-post re-entrant cavity by applying translational symmetry rules and removing all walls between posts. If this is done along one axis only, e.g. $x$, one obtains a one-dimensional lattice. Although this could also be easily achieved in two directions (and maybe three directions if one imbeds posts in a dielectric). In this work we concentrate only on the 1D lattice to simplify the discussion. Side and top views of a rectangular cavity with a chain of multiple posts is shown in Fig.~\ref{meta1}. This represents the first realisation of a multiple post re-entrant cavity and unequivocally show the existence of higher order post modes, which also represents a new way to engineer arrays of coupled microwave oscillators and is a subject of a patent\cite{patent2014}.

At each moment of time, the vector of the electrical field points out of, or into the post. Accordingly, the magnetic field can take one of two possible directions around the post, Clockwise or Anti-Clockwise. These situations are shown in Fig.~\ref{meta1}: Situation "+" describes the case of current flowing up the post and is referred as positive orientation, the opposite orientation of vectors is then understood as the negative direction (or $\pi$ out of phase) with the current flowing down. In the following sections it is demonstrated that the fields of the different posts can exhibit different orientations at the same time depending on the given eigenmode. 

Whereas modes of a single post cavity have already been studied both analytically and experimentally, a system of multiple posts is a new study and contains extra information. That is the relative orientations of the post currents, "$+$" or "$-$" of the corresponding eigenmodes. Combinations of these orientations set a pattern that is specific and unique for each system eigenmode. In the following sections, these patterns of orientations are analysed using 3D electromagnetic field simulation. The most simple kind of multiple post cavity is the double post cavity, which has already been exploited to focus the magnetic field within a small sample. The work demonstrated ultra-strong coupling between photons and magnons within a sub mm YIG sphere situated between the two posts\cite{Goryachev:2014aa}.

Each post of the 1D chain can be understood as a Harmonic Oscillator (HO) in the frequency range where the transverse effects along the $y$ coordinate can be neglected, with an associated capacitance formed by the gap, and inductance formed by the post itself. In this chain configuration two HOs (or posts) are coupled only through the magnetic field since the overlap of electrical field between the posts is negligible in the limit of a small gap. As a result the system of identical posts can be formalised with the following Hamiltonian:
\begin{equation}
\label{H003GH}
\left. \begin{array}{ll}
\displaystyle H = \sum_{i=1}^N\Big[\frac{\phi_i^2}{2L}+\frac{q^2_i}{2C}\Big] - \sum_{i=1}^{N-1}G_{i,i+1}\phi_i\phi_{i+1} \\
\end{array} \right. 
\end{equation}
where $N$ is the total number of posts, $q_i$ and $\phi_i$ are corresponding charge and fluxes, $G_{i,i+1}$  is the coupling between two posts $i$ and $i+1$ set mostly by the distance between posts and can be understood as a mutual inverse inductance. The model is based on an approximation of the re-entrant cavity as an $LC$ circuit\cite{reen0} formed by its capacitive and inductive parts. In addition, Hamiltonian (\ref{H003GH}) considers couplings only between two nearest neighbours. 

\section{Phonon Emulation}

Hamiltonian (\ref{H003GH}) represents an idealised electromagnetic system consisting of identical coupled cavities. The Hamiltonian can be rewritten in the limit of large $N$ and equal couplings $G=G_{i,i+1}$ as follows:
\begin{equation}
\label{H004GH}
\left. \begin{array}{ll}
\displaystyle H = \sum_{i}\Big[\frac{\phi_i^2}{2L^\prime} +\frac{q^2_i}{2C}+\frac{G}{2}(\phi_i-\phi_{i+1})^2\Big] \\
\end{array} \right. 
\end{equation}
where ${L^\prime}^{-1}={L}^{-1}-{G}$. In terms of eigenmodes
\begin{equation}
\label{H005GH}
\left. \begin{array}{ll}
\displaystyle \phi_i = \sum_n \Phi_ne^{ikn},\displaystyle q_i = \sum_n Q_ne^{ikn},\\
\end{array} \right. 
\end{equation}
this Hamiltonian could be rewritten as:
\begin{equation}
\label{H006GH}
\left. \begin{array}{ll}
\displaystyle H = \frac{1}{2C}\sum_k\Big[ C^2\big(\omega_0^2+\omega_k^2\Big) \Phi_k\Phi_{-k}+ {Q_kQ_{-k}}\Big] \\
\end{array} \right. 
\end{equation}
where $\omega_0^2 = \frac{1}{L^\prime C}$ and $\omega_k$ is given by the phonon dispersion relationship:
\begin{equation}
\label{H007GH}
\left. \begin{array}{ll}
\displaystyle \omega_k=\sqrt\frac{G}{C}\sqrt{1-\cos(k)}. \\
\end{array} \right. 
\end{equation}
So, the simple model of the post chain predicts the existence of a phonon-like dispersion curve $\omega_k$ referenced to a bare photon angular frequency $\omega_0$. Building an analogy between electrical and mechanical systems, $Q_k$ plays a role of momentum and $\Phi_k$ of a coordinate whereas $C$ is an analog of the mass. Moreover, taking into account energy $\omega_0$ that is independent of the wavelength, one may draw an analogy with a relativistic particle. Such that in the long wave length limit, the full dispersion relationship could be written as:
\begin{equation}
\label{H007GHff}
\left. \begin{array}{ll}
\displaystyle \omega(k)\approx v^2\sqrt{m^2v^2+k^2}. \\
\end{array} \right. 
\end{equation}
where $v=\sqrt{\frac{G}{C}}$ is an effective "speed of light" and $m=\frac{C}{G^2}\sqrt{\frac{C}{L^\prime}}$ is an effective particle "mass". Accordingly, Hamiltonian (\ref{H006GH}) can be transformed to the Lagrangian for the one-dimensional Klein-Gordon field. 
Existence of such a dispersion curve is verified numerically in the subsequent sections.

\section{Chain of Equidistant Identical Posts}

\begin{figure}[t!]
	\centering
			\includegraphics[width=0.7\textwidth]{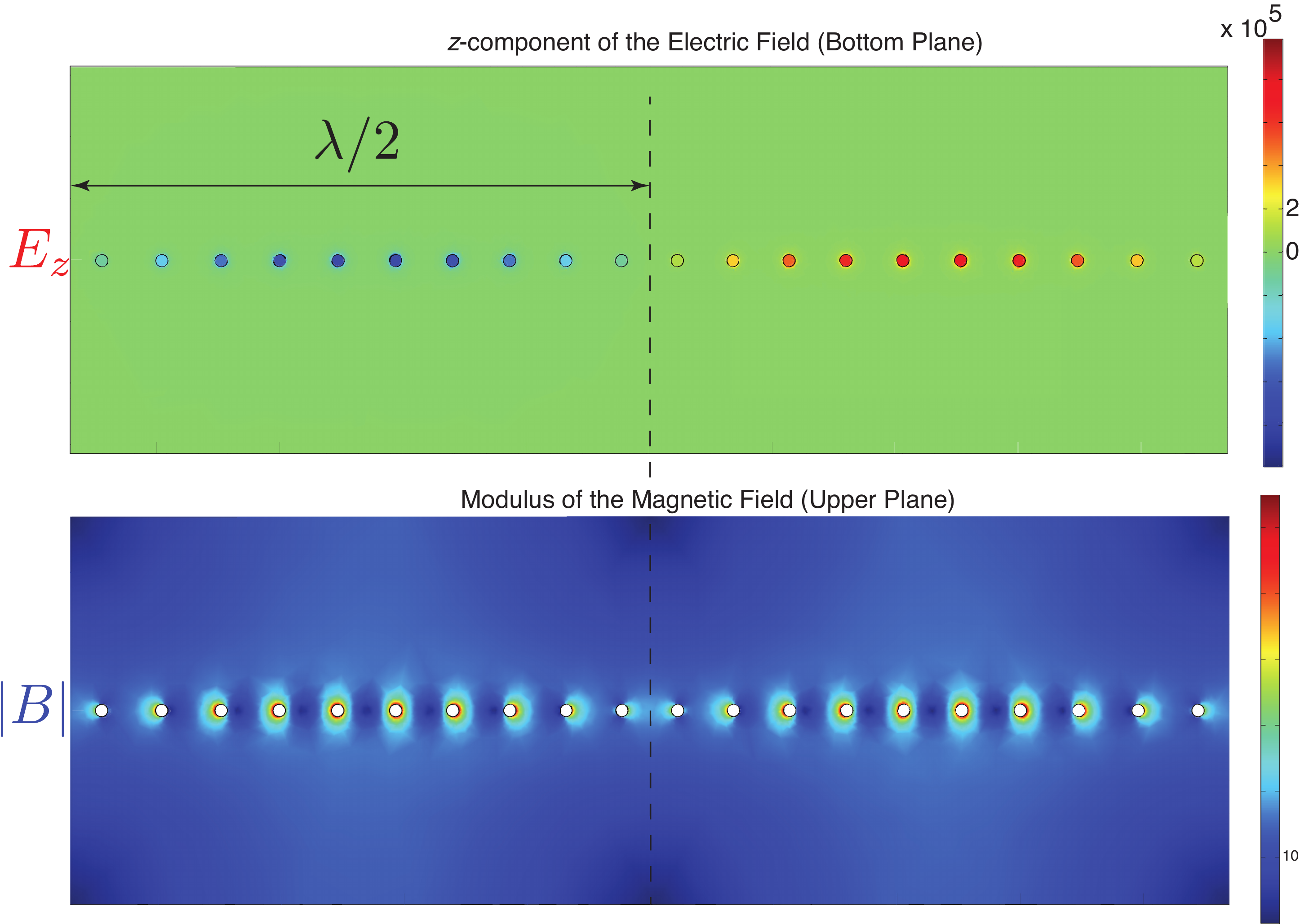}
	\caption{ Electric and magnetic fields at the Bottom and Upper planes for a system mode at $13.0246$~GHz. The two half-waves can be identified by the variation of the field at the posts, which indicates a second order mode $n=2$ of the "acoustic branch".}
	\label{wrA2Ez}
\end{figure}

\begin{figure}[t!]
	\centering
			\includegraphics[width=0.7\textwidth]{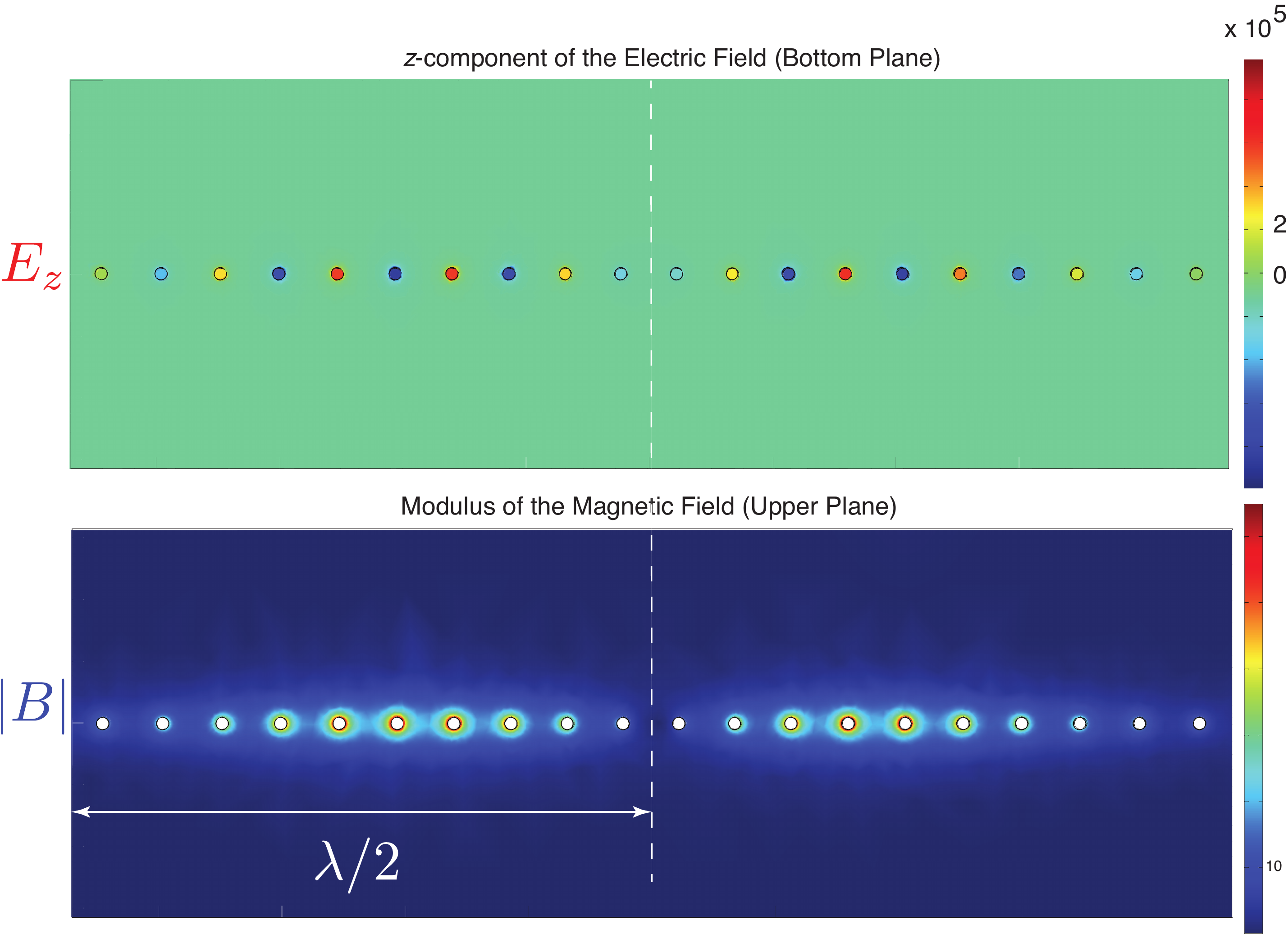}
	\caption{Electric and magnetic fields at the Bottom and Upper planes for a system mode at $29.6247$~GHz. The two half-waves can be identified by the variation of the field at the posts, which indicates a second order mode $n=2$ of the "optical branch".}
	\label{wrB2Ez}
\end{figure}

As a starting point we consider a chain of $N=20$ identical equidistant posts is considered. Simulation results demonstrate the existence of 20 modes with one variation along all dimensions normal to the chain (first order in width). All these modes could be separated into two groups: 1) in-phase currents through the neighbouring posts, which create time-varying regions of posts with positive and negative directions (Fig.~\ref{wrA2Ez}), 2) anti-phase currents through the neighbouring posts creating a time-varying analogue of electrical dipole moments (Fig.~\ref{wrB2Ez}). For each type of mode, one can uniquely identify a mode number $n$ as a number of variations of magnetic and electrical field along the chain, i.e. the number of half-waves. Both Fig.~\ref{wrA2Ez} and Fig.~\ref{wrB2Ez} demonstrate modes with two half-waves.

\begin{figure}[t!]
	\centering
			\includegraphics[width=0.7\textwidth]{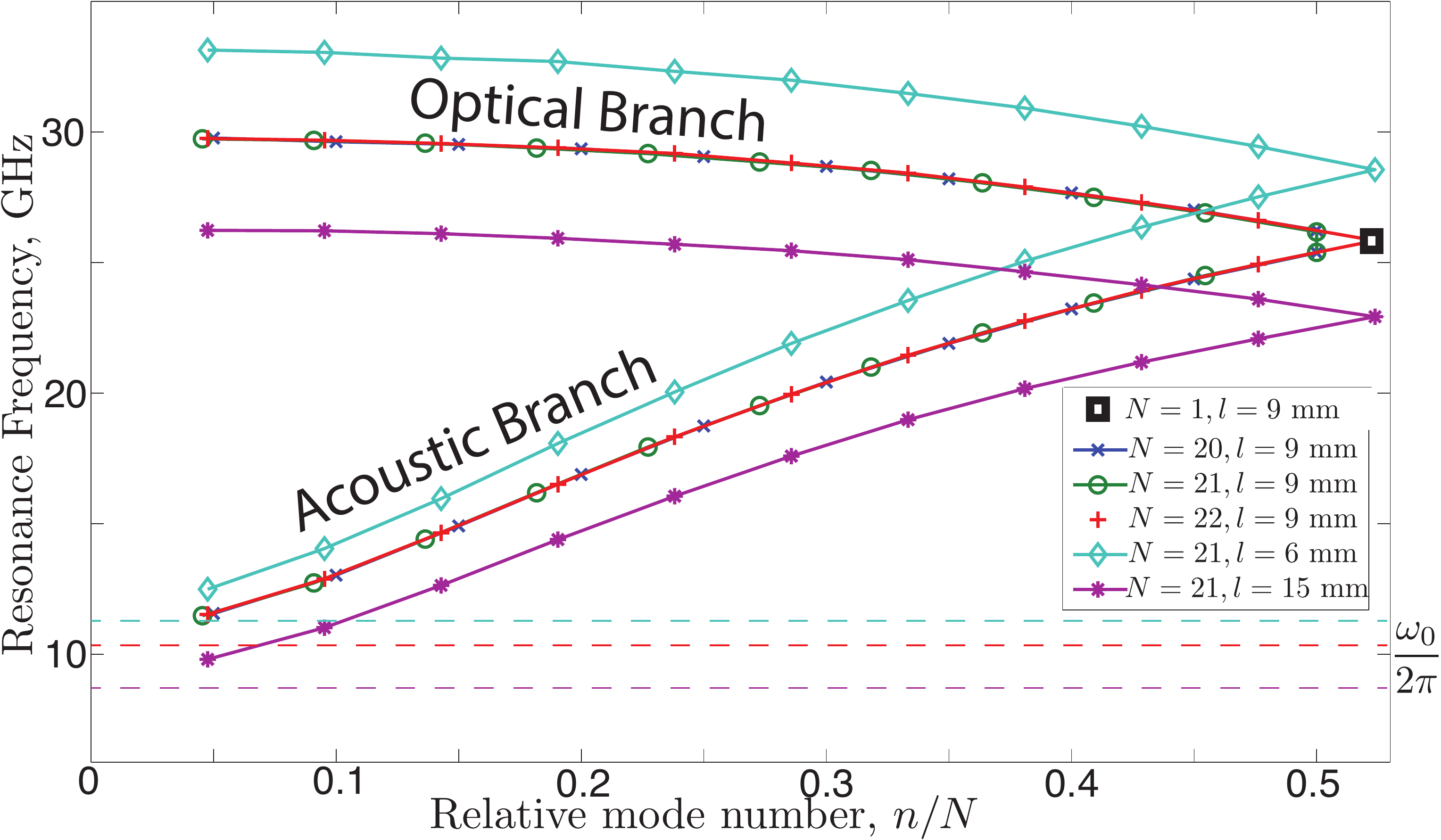}
	\caption{Dispersion curves for a re-entrant chain of posts for different values of $N$ and spacing between posts $h$. The individual square on the far right on the figure shows the case of $N=1$ for $h=9$~mm.}
	\label{dispersion1}
\end{figure}

Resonances of the two types of modes form two distinct types of dispersion curves as shown in Fig.~\ref{dispersion1}. The first type gives an ascending curve where the frequency increases with the normalised wave vector $k = n/N$. On the contrary, the second type demonstrates descending behaviour, i.e. decrease of the frequency with $n/N$. This situations is analogous to acoustic and optical phonons in an one-dimensional lattice. In this interpretation, the role of displacement of an atom is played by the electrical field under the post: Acoustic modes are characterised by the in-phase 'movement' of the neighbouring posts whereas optical modes are characterised by out-of-phase electrical vectors of the nearest neighbours. Although, unlike an optical branch of a mechanical chain of atoms, the acoustical branch in the present system does not demonstrate convergence to zero in the limit of infinitely long wavelengths $k^{-1}$. Instead, it asymptotically approaches $\frac{\omega_0}{2\pi}$ (dashed line in Fig.~\ref{dispersion1}) as the minimal photon energy in (\ref{H006GH}). In the following text, we refer to these branches as acoustic and optical photon dispersion branches which are taken in reference to $\omega_0$. 

Fig.~\ref{dispersion1} shows that all chains with a different number of curves consistently represent the same dispersion curves. The difference of the odd $N$ case is existence of the point where two curves collapse and exhibit the same behaviour. The same frequency correspond to a system with only one post (square in Fig.~\ref{dispersion1}).     

Parameters of the dispersion curves for photons in a chain of posts depend on the post gap, radius and spacing between posts. Fig.~\ref{dispersion1} demonstrate dependence of the curves on the spacing distance. The results demonstrates that the frequencies drop with decreasing density of posts. 

\section{Bandgap}
\label{bg}

\begin{figure}[t!]
	\centering
			\includegraphics[width=0.7\textwidth]{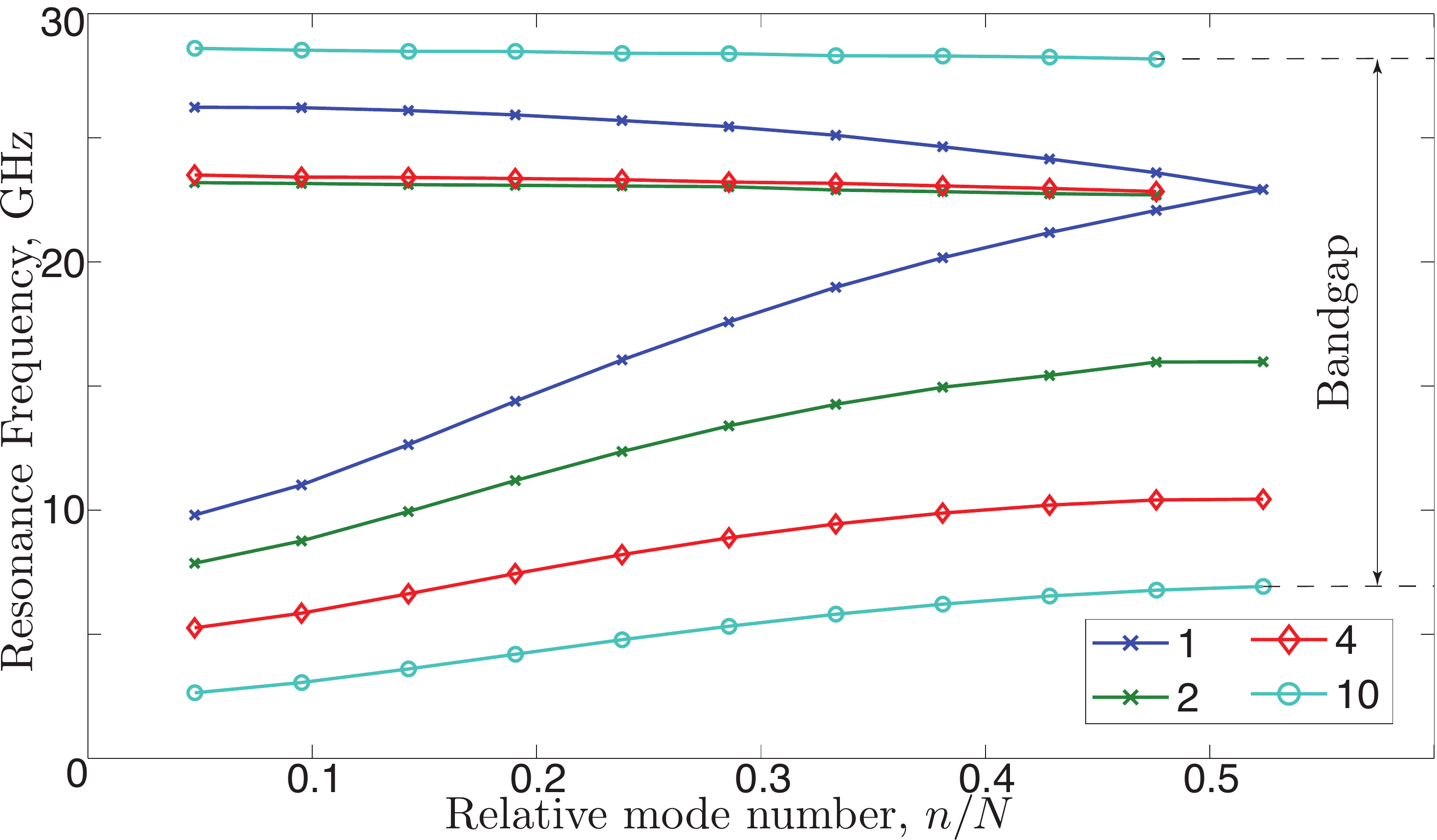}
	\caption{Dispersion curves for a chain of two different post radii, which alternate periodically. The values in the legend (1, 2, 4 and 10) specify ratios between these radii.}
	\label{gap1}
\end{figure}

In a chain of vibrating atoms, a band gap is created by a periodic repetition of atoms with different masses.
Similarly, in the present system a band gap could be created by the variation of post parameters, e.g. its radius or gap. As a result the $L$ and $C$ of each post in the Hamiltonian~(\ref{H003GH}) will depend on its number. The dispersion curves shown in Fig.~\ref{gap1} demonstrate the dispersion characteristic for a chain with periodically changing post radii. As it is clearly seen from the figure, the splitting between two branches increases for increasing ratio between post radii as it is expected from a chain of vibrating atoms. 

\begin{figure}[t!]
	\centering
			\includegraphics[width=0.7\textwidth]{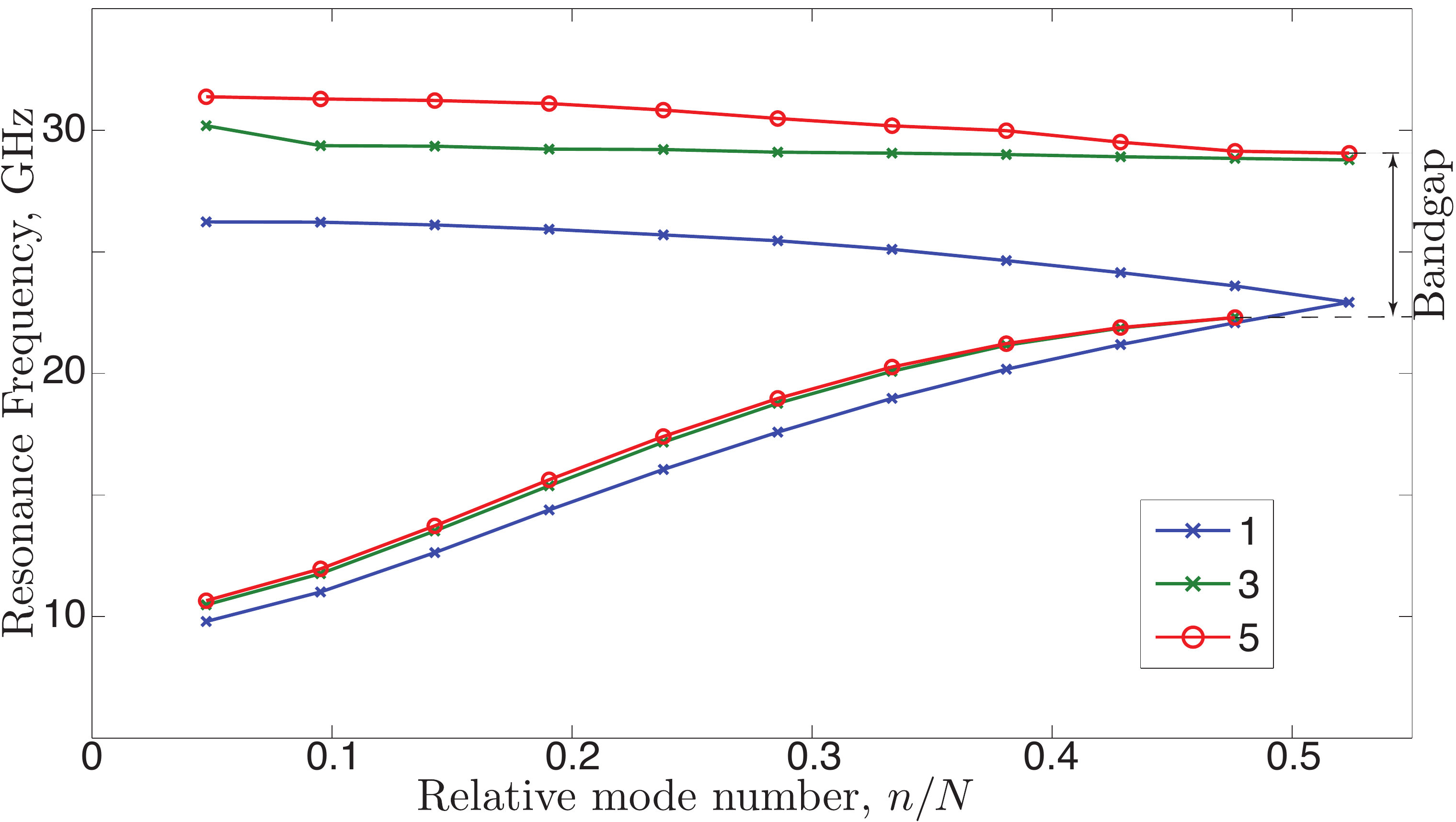}
	\caption{Dispersion curves for a chain of two different post gaps, which alternate periodically. The values in the legend (1, 3 and 5) specify ratios between these gaps.}
	\label{gap2}
\end{figure}

Fig.~\ref{gap2} shows a family of curves for a chain with a periodic change of post gaps. Although a band gap is observed, it shows no further increase after a certain value of ratios. Instead the modes loose their identity, since the re-entrant type of resonance becomes lost.   

\section{Potential Well}

A variation of spacing between the posts could be used to change the energy distribution along the chain. For example, photons can be concentrated at the centre of the chain by increasing gaps between posts in both directions of photon propagation. In this section we consider the case of a potential well, where the post coordinate $x$ depends on the post site number $n$ as follows:
\begin{equation}
\label{H008GH}
\left. \begin{array}{ll}
\displaystyle x_n = \mbox{sign}(n)h/2 +hn + g n^2, \hspace{5pt} n\in\big[-N/2,N/2\big]\\
\end{array} \right. 
\end{equation}
where $h$ and $g$ are some constants. Such a law implies quadratic growth of spacing between posts $G_{i,i+1}$ as one moves from the chain centre. Correspondingly, the coupling energy between the two nearest neighbouring posts decreases. It can be shown that the Hamiltonian density in the limit $N\rightarrow\infty$ and $k\rightarrow 0$ becomes 
\begin{equation}
\label{H004GHi}
\left. \begin{array}{ll}
\displaystyle {H}(x) = \frac{q^2(x)}{2C}+\widetilde{G}(x)(\partial_x\phi(x))^2+\frac{\phi^2(x)}{2L^\prime(x)}, \\
\end{array} \right. 
\end{equation}
where $\widetilde{G}(x)$ grows with $x$ as prescribed by (\ref{H008GH}). As a result, energy of the quasiparticle is minimised at the origin.

\begin{figure}[t!]
	\centering
			\includegraphics[width=0.7\textwidth]{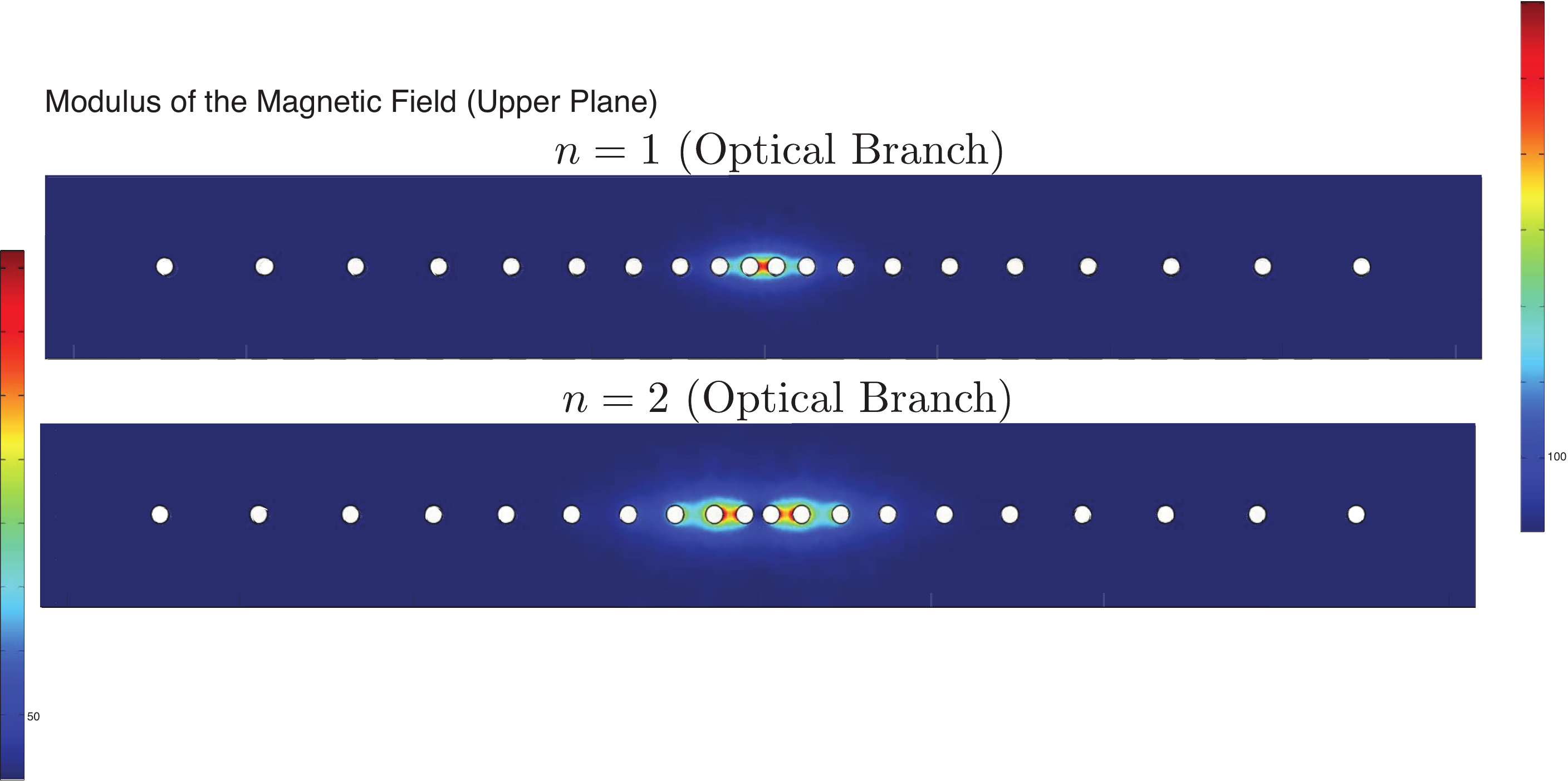}
	\caption{Magnetic field distributions $\big|B\big|$ for two lowest order modes of the optical branch in the case of variable distance between the posts.}
	\label{focn1mag}
\end{figure}

\begin{figure}[t!]
	\centering
			\includegraphics[width=0.7\textwidth]{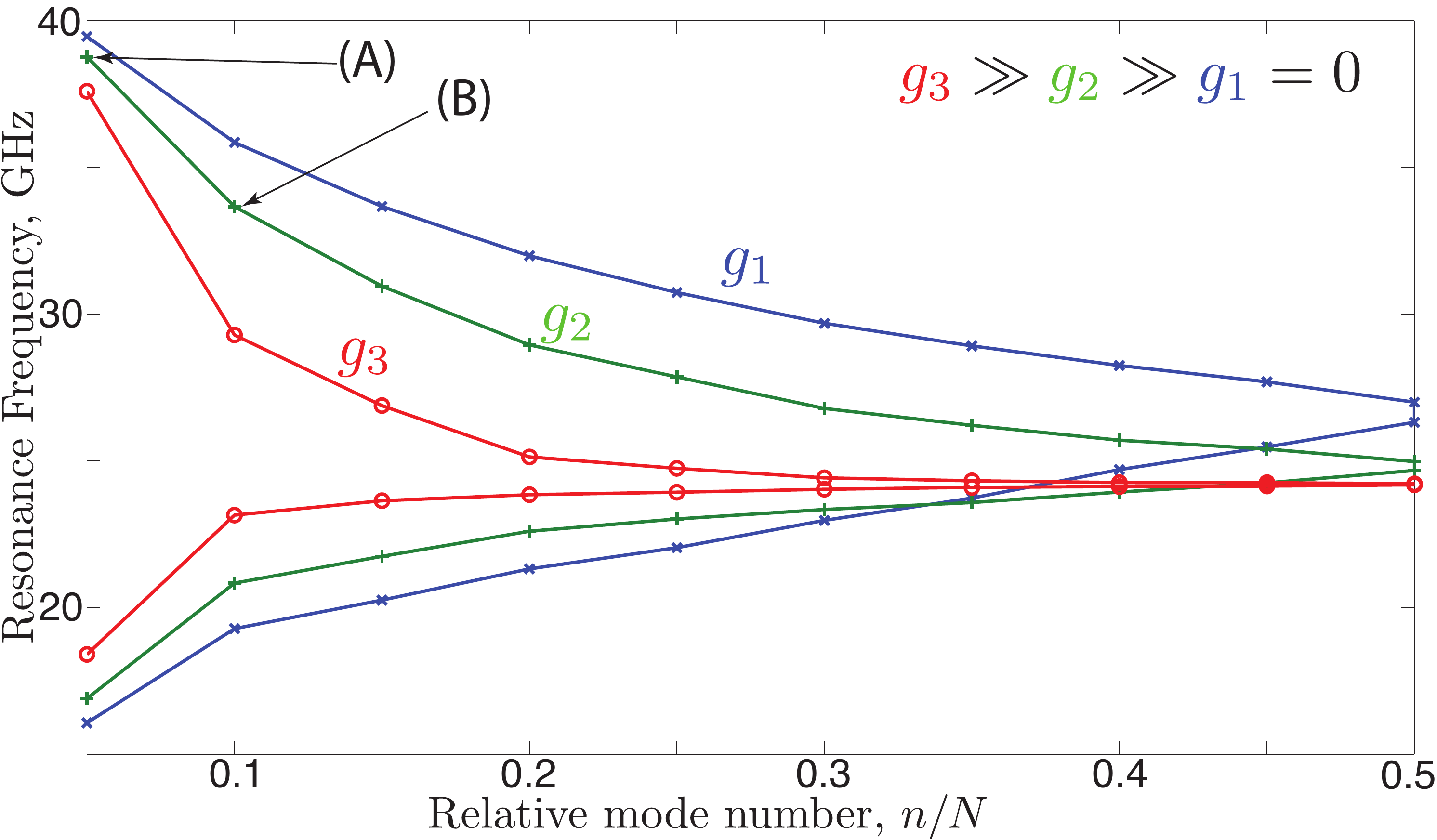}
	\caption{Dispersion curves for a chain of posts with quadratic variation of spacing for different values of the quadratic term constant $g$. (A) and (B) are two cases shown in Fig.~\ref{focn1mag}.}
	\label{focus1}
\end{figure}

The increased concentration of the field is clearly seen in Fig.~\ref{focn1mag}, compared to the uniform post distribution shown in Fig.~\ref{wrB2Ez}. The result could be understood as photon trapping similar to phonon trapping techniques used in high-$Q$ acoustical cavities\cite{Goryachev1,quartzPRL,NatSR}. The variable post separation results in the creation of an effective potential well along the chain, which distorts the dispersion curves (Fig.~\ref{focus1}). It can be inferred from this figure that in the extreme case of $g\rightarrow\infty$, all the modes collapse to the same frequency where all the energy is concentrated between the posts closest to the origin. Similar results can be obtained if the post parameters are quadratically varied along the chain. 

 \section{Lattice Defects}
 \subsection{Interstitial Defect}

\begin{figure}[t!]
	\centering
			\includegraphics[width=0.7\textwidth]{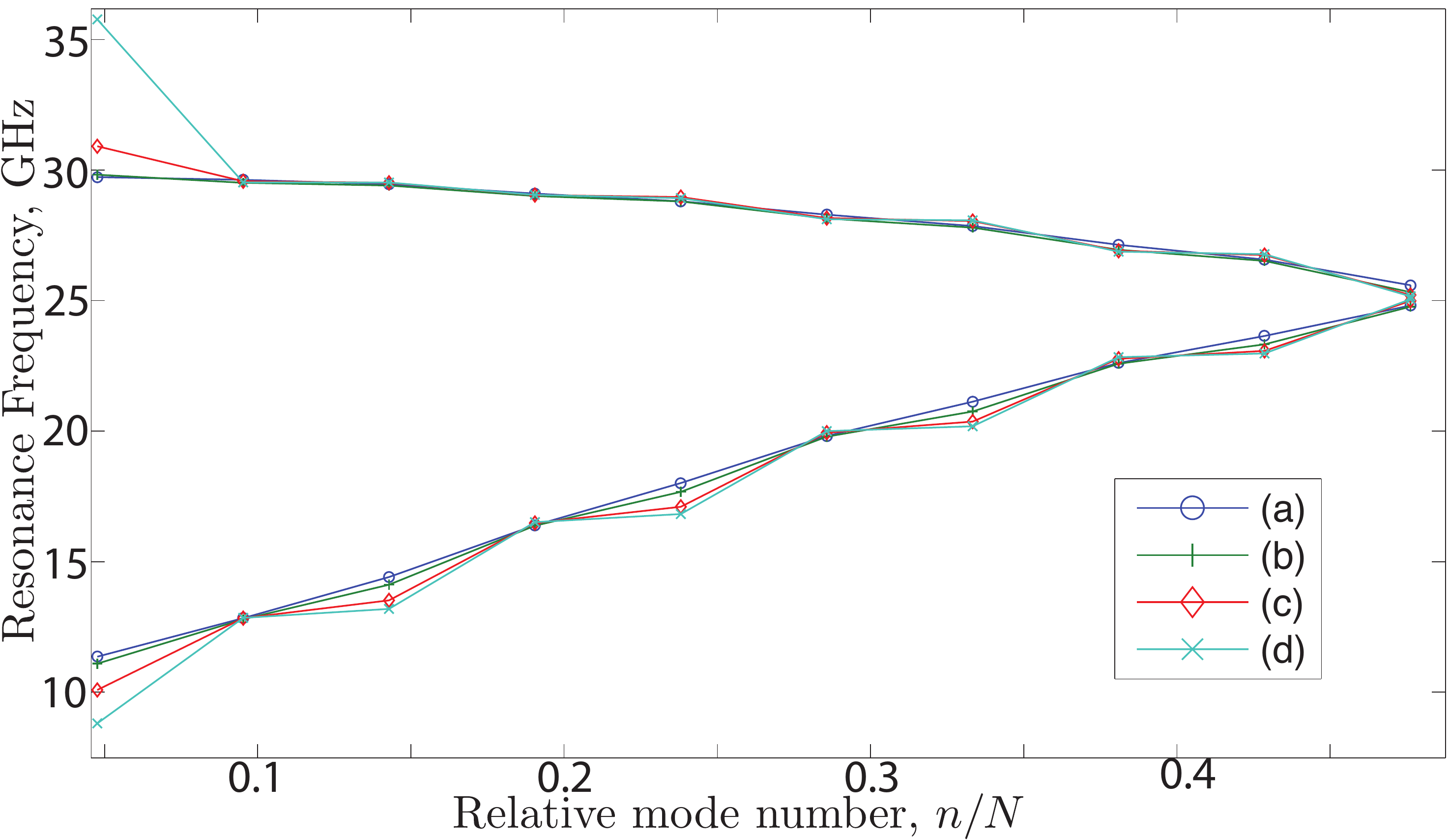}
	\caption{Dispersion curves for a chain of posts with an interstitial impurity in the middle of the chain and for various impurity post radii: (a) $r/4$, (b) $r$, (c) $2r$, (d) $3r$ where $r$ is the chain post radius.}
	\label{impurity1}
\end{figure}

Real solids are always subject to various defects and imperfections of the crystal structure. An example of such a defect is an interstitial impurity. It arises when an alien ion or atom exists in an interatomic space of the crystal. This situation is very common for natural quartz where interstitial $H^+$, $Li^+$, $Na^+$ ions exists as charge compensators near substitutional Al$^{3+}$.
 In the case of the simple chain of atoms, the interstitial impurity may be modelled as an extra post between regular posts of a chain. Simulated dispersion curves for such a system are shown in Fig.~\ref{impurity1}. Various curves are shown for different relative sizes of the impurity post.

Fig.~\ref{impurity1} demonstrates an effect of separation of the chain into two separate chains. In the large impurity limit, the impurity works as a wall (large inertia)  making two separate resonant systems. For this reason, two adjacent points on the dispersion curve tend to have the same angular frequency. One of these frequencies correspond to the left part of the chain and another to the right half-chain. The other limiting case is when the impurity (or post) is infinitesimally small, in this case the system is unperturbed. Thus, in this limit the curves converge to the dispersion curve of the ideal chain.

 \subsection{Substitutional Defect}
 
\begin{figure}[t!]
	\centering
			\includegraphics[width=0.7\textwidth]{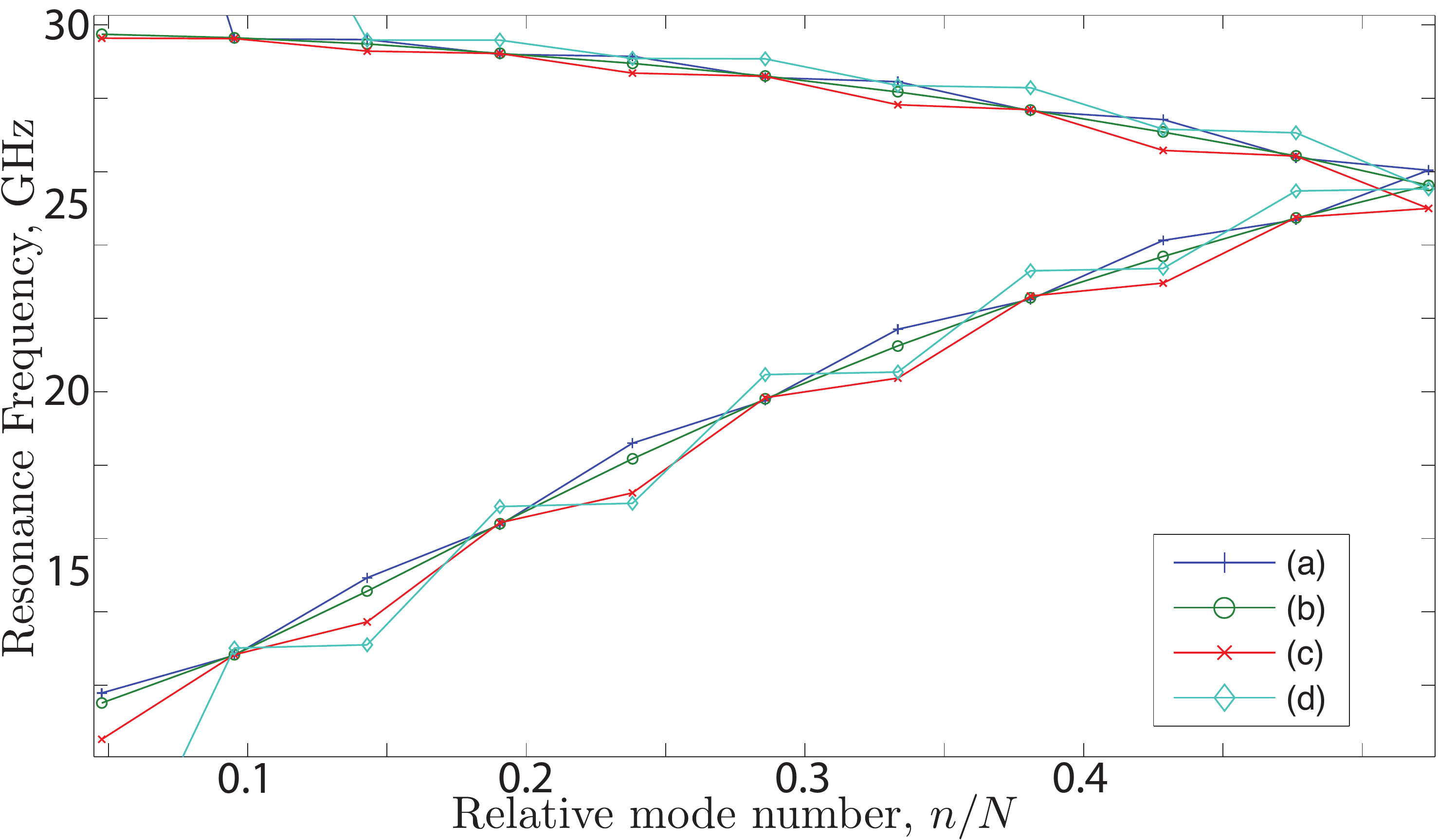}
	\caption{Dispersion curves for a chain of posts with an substitutional impurity in the middle of the chain and for various impurity post radii: (a) $r/7$, (b) $r$, (c) $2r$, (d) $7.5r$ where $r$ is the chain post radius.}
	\label{impurity2}
\end{figure}

 Another common type of impurity in a crystal is a substitutional impurity. The situations arises when an alien atom substitutes a normal atom in a crystal. This situation also can be simulated with a chain of posts with the results shown in Fig.~\ref{impurity2}.

The effect of a large substitutional impurity is similar to that of a large interstitial impurity. The substitutional post acts as a wall separating the lattice of resonators into two sub-chains. So, in the limit of an infinitely large impurity, two subsystems are completely separated and uncoupled giving two separate dispersion curved. But unlike in the case of infinitesimally small limit of the substitutional impurity, an infinitesimally small substitutional defect effectively acts as a gap. This gap is also a kind of separator that decreases coupling between two ideal sub-chains. Thus, a similar effect of doubling the dispersion curves is also observed in this limit (Fig.~\ref{impurity2}).  
 
  \section{Quasicrystal}

\begin{figure}[t!]
	\centering
			\includegraphics[width=0.7\textwidth]{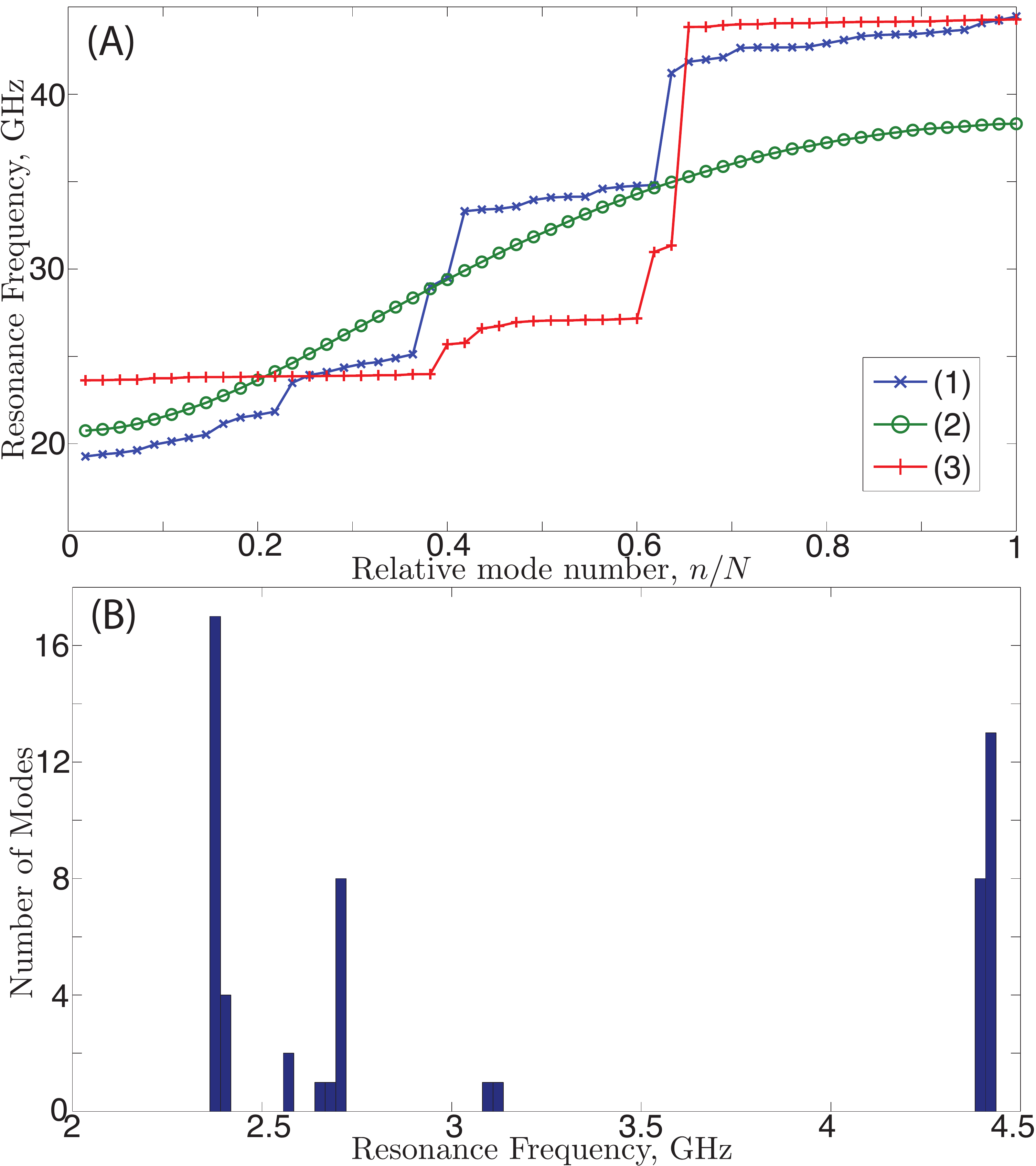}
	\caption{Eigenmode distribution for a Fibonacci chain of posts: (A) dispersion relation, (B) histogram of eigenfrequencies. (1) and (3) are two different realisations of a chain, (2) is a regular chain with one type of a post.}
	\label{quasi1}
\end{figure}

Quasicrystals\cite{quasi} are an ordered form of matter that do not have translational symmetry and a lattice of posts can also be used to study both 1D and 2D quasicrystal structures. One of the simplest examples of quasicrystals is a Fibonacci chain formed by the rule $F_n = F_{n-1}+F_{n-2}$. Such a chain does not have translational symmetry unlike the regular chain of two posts studied in Section~\ref{bg}. For example, for a chain of $N=21$ posts the structure is $LSLLSLSLLSLLSLSLLSLSL$, where $L$ and $S$ represent posts of two different types or two different distances between two neighbouring posts. Fig.~\ref{quasi1} compares dispersion characteristics of such metamaterial with the regular one for $N=33$. One can see from these results that quasicrystalline metamaterials demonstrates effects significantly different from that of regular chains. In particular, multiple discontinuities of spectra are observed. 

\section{Possible Applications} 

The proposed type of the metamaterial may be used for various applications in physics. Its main advantage is high tunability\cite{sensiv} and potentially higher quality factors than the 2D equivalent. High tunability of the order of $1$~GHz/$\mu$m for $\mu$m-size gaps has been achieved due to the concentration of the electrical field in the tiny gaps\cite{reen1}. Such high tunability has been utilised for transducers for gravitational wave antennae\cite{pimentel2008}. Also, high quality factors $Q>10^8$ have been demonstrated for superconducting re-entrant-type resonators\cite{Bassan2008}. These 3D closed cavities out perform all known 2D structures popular in Circuit Quantum Electrodynamics (cQED)\cite{cqed2} with the reported value only bettered by cryogenic dielectric resonators at low temperatures\cite{Creedon:2011wk} and the best acoustical systems\cite{quartzPRL,NatSR}. 

The functionality of the linear 3D split-ring metamaterial can be extended by the addition of nonlinear or parametric elements:
For example, the 3D split ring cavity lattice could be used for cQED experiments by incorporating Josephson Junctions. In this case, a Josephson Junction can be formed by the dielectric interface between the post and the cavity bottom. Such pin contact junctions have been investigated at the very beginning of the era of nonlinear superconducting systems\cite{Silver:1967to,Zimmerman:1970ad}. The Hamiltonian of this system can be written in the following form:
\begin{equation}
\label{H008GHf}
\left. \begin{array}{ll}
\displaystyle H = \sum_{i=1}^N\Big[\frac{\phi_i^2}{2L}+\frac{q^2_i}{2C}-E_J\cos\phi_i\Big] - \sum_{i=1}^{N-1}G\phi_i\phi_{i+1}, \\
\end{array} \right. 
\end{equation}
where $E_J$ is the Josephson energy. {This system can be interesting from the point of view of emulation of new phases of matter (which has gathered momentum in recent years)\cite{1367-2630-12-9-093031,PhysRevLett.110.163605}. 
In particular, a} chain of such posts could serve as an alternative realisation of the supersolid matter state\cite{Kim:2004so}. { Due to the high quality factors of the structure, the resulting nonlinearity may be apparent at a single photon level. This effect could be used to generate non-classical states of microwave radiation via the implementation of a Kerr-medium\cite{Kirchmair:2013aa} with significantly larger coherence times (or lower artificial medium losses). Although the confined structure of the cavity makes external control of the Josephson junctions difficult, one can introduce superconducting biasing coils in the same manner as is achieved for signal loop antennas.}

High sensitivity to mechanical motion of the gap makes the re-entrant cavity an excellent transducer of mechanical motion. In principle, the chain in this case may be thought as an array of inter-coupled harmonic oscillators controlled by cavity photon modes:
\begin{equation}
\label{H009GH}
\left. \begin{array}{ll}
\displaystyle H = \sum_{i=1}^N\hbar\Big[(\omega + k(a_i+a_i^\dagger)) b_i^\dagger b_i+\omega_ma_i^\dagger a_i\Big] -\\
\displaystyle- \sum_{i=1}^{N-1}\widetilde{G}(b^\dagger_ib_{i+1}+bb_{i+1}^\dagger). \\
\end{array} \right. 
\end{equation}
where $\omega_m$, $a_i^\dagger (a_i)$ are the angular frequency and the creation (annihilation) operators of the mechanical resonators under each post, $\omega$, $b_i^\dagger (b_i)$ are the angular frequency and the creation (annihilation) operators of the electromagnetic resonance associated with each post. 

Concentration of the magnetic field around and between certain posts also allows the coupling of selected photon modes to spins states of the crystal impurities embedded in the resonator, which may be used to achieve photon-spin strong coupling regimes\cite{Probst:2013zg}. For example, recently the ultra-strong coupling regimes between magnons and a two post photon cavity has been demonstrated\cite{Goryachev:2014aa}. This may be easily extended to multiple post and spin systems.

{ In summary, the proposed structure can be used for the implementation of more complex Hybrid Quantum Systems (HQS)\cite{PhysRevLett.92.247902,PhysRevLett.102.083602,PhysRevLett.103.043603,Xiang:2013aa}. For example, the proposed lattice of 3D cavities can integrate microwave photon cavities with optomechanics\cite{Kippen}, superfluid optomechanics\cite{superf}, superconducting qubits\cite{PhysRevB.86.100506,PhysRevLett.107.240501} as well as spins-in-solids\cite{PhysRevB.90.100404,Goryachev:2014aa,YSO1}.}

Further applications for the presented metamaterial could also include (but are not limited to)\cite{patent2014}:
\begin{itemize}
\item Spectroscopy of small samples of magnetic materials, which requires maximization of the magnetic filling factor of the material under study\cite{Goryachev:2014aa}. Advantages include high tunability by changing the electrical length by the gap by either mechanical or dielectric methods, and the ability to focus the magnetic field between posts of close proximity. Also frequencies much smaller than the natural resonant frequencies of a small sample may be realized. The same set up can be used to characterize the magnetic properties of a material without the influence of dielectric effects due to the separability of the electric and magnetic fields.

\item Implementation of arrays, which magnetically couple to more than one sample simultaneously. Samples could include spins or arrays of qubits. This would allow complex qubit/spin circuitry, with engineered couplings to specified samples for quantum computing and engineering applications. For example, if each post is associated with a Josephson Junction and cooled to low enough temperatures, the multi-post lattice structure provides a new way of arrangement of quantum information, including 2D and 3D arrangement of qubits. The different modes of the structure can then be used to address and probe different subsystems.

\item The study of electrical response of dielectric materials (materials characterization); some applications could require either concentration of the electric field or its gradient, this could be achieved by putting a sample (film or wafer) under different posts, thus different modes will apply different electric field patterns to the sample.

\item Highly tuneable, multi-mode filters are realisable due to the existence of several modes of the multi-post lattice cavity. Such systems have application in multi-band antenna applications or as highly tuneable notch filters for communication jamming applications. Unlike previous inventions, we propose to utilise all the resonant mode frequencies of the structure, which is equal to the number of posts. 

\item Space-resolving displacement detectors May be also realised, as a single post re-entrant cavity is a highly sensitive displacement sensor. The multi-post structure has a few resonant frequencies, with each of them with a unique sensitivity distribution in the gap plane. Thus such a structure will allow the resolution of the displacement patterns by identifying frequency shifts of all resonant modes.

\item Multi-mode gas/fluid sensor could be realised as an extension of a single post cavity, allowing multi-frequency analysis due to existence of several modes to probe the gas/fluid at different frequencies. 
\item Band gap filters and isolators could be realised using a multi-post metamaterial with posts of different size/gaps/spacings, leading to the creation of wide band gaps for isolation purposes. Existence of the such gaps is demonstrated in the present paper. In addition to that, the proposed type of metamaterial is a convenient tool to study novel structures of solids such as quasicrystals in one and two dimensions, which is not possible in real matter.

\end{itemize}

{ In summary, the work proposes a new base for creating artificial materials based on a 3D cavity structure known as re-entrant (or Klystron) cavity. The system has several advantages over existing counterparts: 1) better field confinement results in $Q$-factors superior to that of 2D elements, 2) direct mechanical tunability with extraordinary tuning coefficients that is harder to implement in 2D structures, 3) high concentration of electric and magnetic fields that can be engineered to exhibit higher filling factors than in planar devices, 4) spatial separation of the electrical and magnetic fields, 5) the closed cavity makes possible applications with gases and fluids. Moreover, unlike other systems, the proposed concept is ideal for implementation of HQS due to possibility to incorporate various elements exhibiting quantum properties. On the other hand, the multi-post lattice structure is advantageous over its one-post counterpart due to existence of multiple resonances. This feature can be used for spectroscopy purposes that probes different frequencies, as multi-resonance filters or to increase overall sensitivity to mechanical motion. }

\section*{Acknowledgments}
This work was supported by the Australian Research Council Grant No. CE110001013.

\hspace{15pt}

\section*{References}

\providecommand{\newblock}{}

\end{document}